\documentclass[a4paper,12pt]{article}
\pdfoutput=1
\usepackage{cite}
\usepackage{amsmath}
\usepackage{amsfonts}
\usepackage{amssymb}
\usepackage{graphicx, rotating}
\usepackage{epsfig}
\usepackage{latexsym}
\usepackage{graphicx}
\usepackage{color}
\usepackage{amsmath,bm,amssymb}

\usepackage{color}
\usepackage[english]{babel}
\usepackage{hyperref}

\newcommand{\Slash}[1]{{\ooalign{\hfil#1\hfil\crcr\raise.167ex\hbox{/}}}}

\newcommand{\beq}{\begin{equation}}  \newcommand{\eeq}{\end{equation}}
\newcommand{\bef}{\begin{figure}}  \newcommand{\eef}{\end{figure}}
\newcommand{\bec}{\begin{center}}  \newcommand{\eec}{\end{center}}
\newcommand{\non}{\nonumber}  
\newcommand{\laq}[1]{\label{eq:#1}}  

\newcommand{\Eq}[1]{Eq.~(\ref{eq:#1})}
\newcommand{\Eqs}[1]{Eqs.~(\ref{eq:#1})}
\newcommand{\eq}[1]{(\ref{eq:#1})}

\newcommand{\vev}[1]{ \left\langle {#1} \right\rangle }

\newcommand{\SU}[1]{{\rm SU{#1} } }

\def\({\left(}
\def\){\right)}

\def\O{\mathcal{O}}
\def\U{\mathop{\rm U}}

\newcommand{\AND}{~{\rm and}~}
\newcommand{\EV}{ {\rm \, eV} }

\newcommand{\MEV}{ {\rm \, MeV} }
\newcommand{\GEV}{ {\rm \, GeV} }

\def\a{\alpha}

\def\d{\delta}

\def\f{\phi}
\def\g{\gamma}
\def\h{\theta}
\def\k{\kappa}
\def\l{\lambda}
\def\m{\mu}

\def\p{\psi}

\def\x{\xi}

\def\D{\Delta}
\def\G{\Gamma}

\def\L{\Lambda}
\def\F{\Phi}

\def\ol{\overline}
\def\tl{\tilde}
\def\*{\dagger}

\begin{document}
\renewcommand\bibname{\Large References}

\begin{center}

\hfill   TU-1115\\
\hfill  IPMU20-0130\\

\vspace{1.5cm}

{\Large\bf  Kilobyte Cosmic Birefringence from ALP domain walls}\\
\vspace{1.5cm}

{\bf  Fuminobu Takahashi$^{1,2}$, Wen Yin$^{3}$}

\vspace{12pt}
\vspace{1.5cm}
{\em 
$^{1}$Department of Physics, Tohoku University,  
Sendai, Miyagi 980-8578, Japan \\
$^{2}$Kavli Institute for the Physics and Mathematics of the Universe (WPI),
University of Tokyo, Kashiwa 277--8583, Japan\\
$^{3}${ Department of Physics, Faculty of Science, The University of Tokyo,  \\ 
Bunkyo-ku, Tokyo 113-0033, Japan} \vspace{5pt}}

\vspace{1.5cm}
\abstract{
ALP domain walls without strings may be formed in the early Universe. We point out that
such ALP domain walls lead to both isotropic and anisotropic birefringence 
of cosmic microwave background (CMB) polarization, which reflects spatial configuration of the domain walls at the recombination. 
The polarization plane of the CMB photon coming from each domain is either not rotated at all or rotated by a fixed angle. For domain walls following the scaling solution, 
the cosmic birefringence of CMB is characterized by
 $2^{N}$, i.e. $N$-bit, of information with $N = {\cal O}(10^{3-4})$ 
 being equal to the number of domains at the last scattering surface,
and thus the name,  {\it kilobyte cosmic birefringence}. 
 The magnitude of the isotropic birefringence is consistent with the recently reported value, while the anisotropic one is determined by the structure of domains at the last scattering surface. The predicted cosmic birefringence is universal over a wide range of the ALP mass and coupling to photons. The detection
 of both signals will be a smoking-gun evidence for the ALP domain walls without strings.
}

\end{center}
\clearpage

\setcounter{page}{1}
\setcounter{footnote}{0}

\section{Introduction}

Light axions may be ubiquitous in nature. In the string or M theory, there often appear many axions, which we collectively denote by $\phi$, and their cosmological and phenomenological implications have been studied in a context of e.g. the axiverse or 
axion landscape scenarios \cite{Witten:1984dg, Svrcek:2006yi,Conlon:2006tq,Arvanitaki:2009fg,Acharya:2010zx, Higaki:2011me, 
Cicoli:2012sz,Demirtas:2018akl}. The axion enjoys a discrete shift symmetry,
\begin{eqnarray}
\laq{shift}
\phi  \rightarrow \phi + 2  \pi f_\f ~,
\end{eqnarray}
where $f_\phi$ is the decay constant of the axion.  Also, such axions  appear through the spontaneous symmetry breaking of a 
global U(1) Peccei-Quinn (PQ) symmetry~\cite{Peccei:1977hh,Peccei:1977ur,Weinberg:1977ma,Wilczek:1977pj}. 
The axion potential is usually generated via non-perturbative effects and its mass can be exponentially suppressed.
One unique prediction of such theories with axions is the existence of degenerate vacua. If the degenerate  vacua are populated
in space, they are separated by domain walls. 

The axion may have a coupling to photons via anomaly,
\begin{align}
\label{eq:int}
{\cal L} & = c_\g \frac{\a}{4 \pi} \frac{\phi}{f_\f} F_{\mu \nu} \tilde F^{\mu \nu} 
\equiv \frac{1}{4} g_{\phi\g\g} \phi F_{\mu \nu} \tilde F^{\mu \nu},
\end{align}
 where $c_\g$ is an anomaly coefficient,  $\a$ the fine structure constant, $F_{\mu \nu}$ and  $\tl{F}_{\mu \nu}$ the field strength and its dual,
 respectively.
The axion with a coupling to photons is also called  an axion-like-particle (ALP) in the literature. 
See Refs.~\cite{Jaeckel:2010ni,Ringwald:2012hr,Arias:2012az,Graham:2015ouw,Marsh:2015xka, Irastorza:2018dyq, DiLuzio:2020wdo} for a review on axions and related topics. 
 
Recently, a hint of the isotropic cosmic birefringence (CB) of cosmic microwave background (CMB) polarization was reported with the
rotation angle~\cite{Minami:2020odp},
 \beq
 \laq{measure}
 \beta= 0.35 \pm 0.14 {\rm~ deg },
 \eeq
based on the re-analysis of the Planck 2018 polarization data using a novel method~\cite{Minami:2019ruj,Minami:2020xfg,Minami:2020fin}. 
One plausible mechanism to induce the 
CB is to introduce a temporally varying and/or spatially non-uniform 
ALP~\cite{Carroll:1989vb, Carroll:1991zs, Harari:1992ea,Carroll:1998zi,Lue:1998mq,Pospelov:2008gg,Fedderke:2019ajk,Agrawal:2019lkr,Fujita:2020aqt}. In particular, quantum fluctuations generated during inflation have often been considered as the origin 
of spatial inhomogeneity of the ALP, in which case the scale-invariant anisotropic CB is predicted. 
The interpretation {of the hint for isotropic CB} along these lines was  discussed in Refs.~\cite{Minami:2020odp, Fujita:2020ecn}.\footnote{It was also pointed out that,
in a flat-top and flat-bottomed potential, the $H_0$ tension can be relaxed~\cite{Fujita:2020ecn}. Such a potential  had been 
studied in a context of multi-natural inflation~\cite{Czerny:2014wza,Czerny:2014qqa,Croon:2014dma,Higaki:2015kta}.
The unification of inflaton and dark matter in terms of the ALP was also studied using the same kind of potential~\cite{Daido:2017wwb,Daido:2017tbr,Takahashi:2019qmh}. }.
In this paper, we propose a different scenario using the ALP domain walls to induce both isotropic and anisotropic CB. In particular, the predicted isotropic CB nicely explains the reported rotation angle \eq{measure} over a wide range of the ALP mass and coupling to photons.

When a photon travels in a slowly-varying ALP  background,
the polarization angle $\Phi$ changes following  \cite{Carroll:1989vb,Carroll:1991zs,Harari:1992ea}
\beq
\laq{delF}
\dot{\F}\approx \frac{ c_\g \a}{2\pi}\frac{{ \hat{r}^\mu \partial_\m \f}}{f_\f},
\eeq
where $\hat{r}^\mu$ is a normalized photon 
four-momentum;
e.g. $\hat{r}=(1,0,0,1)$ when the photon travels in the positive $z$ direction. Integrating the above equation along the line of sight  from the last scattering surface (LSS) to us, we obtain a net rotation of the polarization plane,
 \beq
 \D\F(\Omega) = 
  0.42 {\rm~ deg } \times c_\g \( \frac{\f_{\rm today}-\f_{\rm LSS}(\Omega) }{2\pi f_\phi}\),
 \eeq
 {where $\f_{\rm today}$ and $\f_{\rm LSS}(\Omega)$  are the axion field value at the solar system today, and at the LSS, respectively, and $\Omega$ denotes the angular direction specified by polar coordinates $(\theta,\varphi)$.}
 The isotropic CB is obtained by
 \beq
 \beta= \frac{1}{4\pi} \int{ d\Omega \,{\D \F}(\Omega)}.
 \eeq
Interestingly, if the change of the axion field value is equal to the shift of the discrete shift symmetry transformation \eq{shift}, i.e., $\f_{\rm today}-\f_{\rm LSS}(\Omega) =2\pi f_\phi$,
the rotation angle $\beta$ is given by $\beta = 0.42\, c_\gamma \,{\rm deg}$, which is {intriguingly} close to the observed value~\eq{measure} if $c_\gamma = {\cal O}(1)$. 
This coincidence has led us to study the CB induced by the axion domain walls separating two adjacent vacua. As we will see, the axion domain walls induce both isotropic and anisotropic CB.

In this paper, we show that ALP domain walls without strings can be naturally produced in the early universe.
Once formed,  domain walls are considered to follow the scaling solution~\cite{Press:1989yh}. If such ALP domain walls are formed  before recombination, 
there {will be} ${\cal O}(10^{3-4})$ domains on the LSS with sizes of the order of the Hubble horizon.
Depending on which vacuum the axion resides in each domain, 
the rotation angle $\D \F(\Omega)$ takes 
one of the two possible values,  either zero or $0.42 c_\gamma$ deg.
Thus, the CB from each domain
carries one-bit information, thus the name {\it the kilobyte cosmic birefringence} {(KBCB)}.
The KBCB consists of both isotropic and anisotropic contributions.
The former is consistent with \eq{measure} if $c_\gamma = {\cal O}(1)$, and the latter is expected to have a peculiar {pattern} which reflects the configuration of domain walls on the LSS. 
We also discuss phenomenological implications of our scenario. 

Before closing the introduction, let us comment on the related works in the past. The anisotropic CB induced by axionic strings and domain walls  attached to them was studied in detail in  Ref.~\cite{Agrawal:2019lkr} where they showed that
the induced anisotropic CB retains the information on the fine-structure constant times the anomaly coefficient. One of the differences of the present work from Ref.~\cite{Agrawal:2019lkr} is that we focus on domain walls without strings, which induce both isotropic  and anisotropic CB.
The absence of strings makes the net rotation angle of the CMB polarization take one of the fixed values, depending on which vacua the axion resides on the LSS. This is because contributions from the domain walls along the line of sight are canceled out, and what we observe today directly reflects the information on the LSS. This should be contrasted to the case with strings where the anisotropic CB receives random contributions from strings along the line of sight. 
These features allow our scenario to predict the CB of a very unique nature, which can be tested by future observations. Furthermore, unlike the model with uniform scalar field motion, the predicted CB is universal over a wide range of the ALP mass and coupling to photons. This is due to the scaling behavior of domain walls.
Moreover, our scenario allows for heavier axion masses and the parameter space with a relatively small axion decay constant, {a part of} which can be probed by future gamma-ray observations.

The structure of this paper is organized as follows. 
In the next section, we show the mechanism of the domain wall formation without strings and discuss the conditions for the mechanism to work.
In Sec.\ref{sec:KB}, the KBCB and its phenomenology are discussed. {In particular, we estimate the predicted angular power spectrum of the anisotropic CB based on a simple model of the domain-wall network.}
The last section is devoted to discussion and conclusions.

\section{Formation of domain walls  without strings}\label{sec:2} 
\subsection{Mechanism}
Here we show that domain walls can be formed without strings by using a simple model, and clarify the condition for the 
mechanism to work. Later in this section, we consider two models with a negative Hubble-induced mass term and a mixing with the QCD axion, in which the condition
is naturally realized. 

\begin{figure}[!t]
\begin{center}  
   \includegraphics[width=145mm]{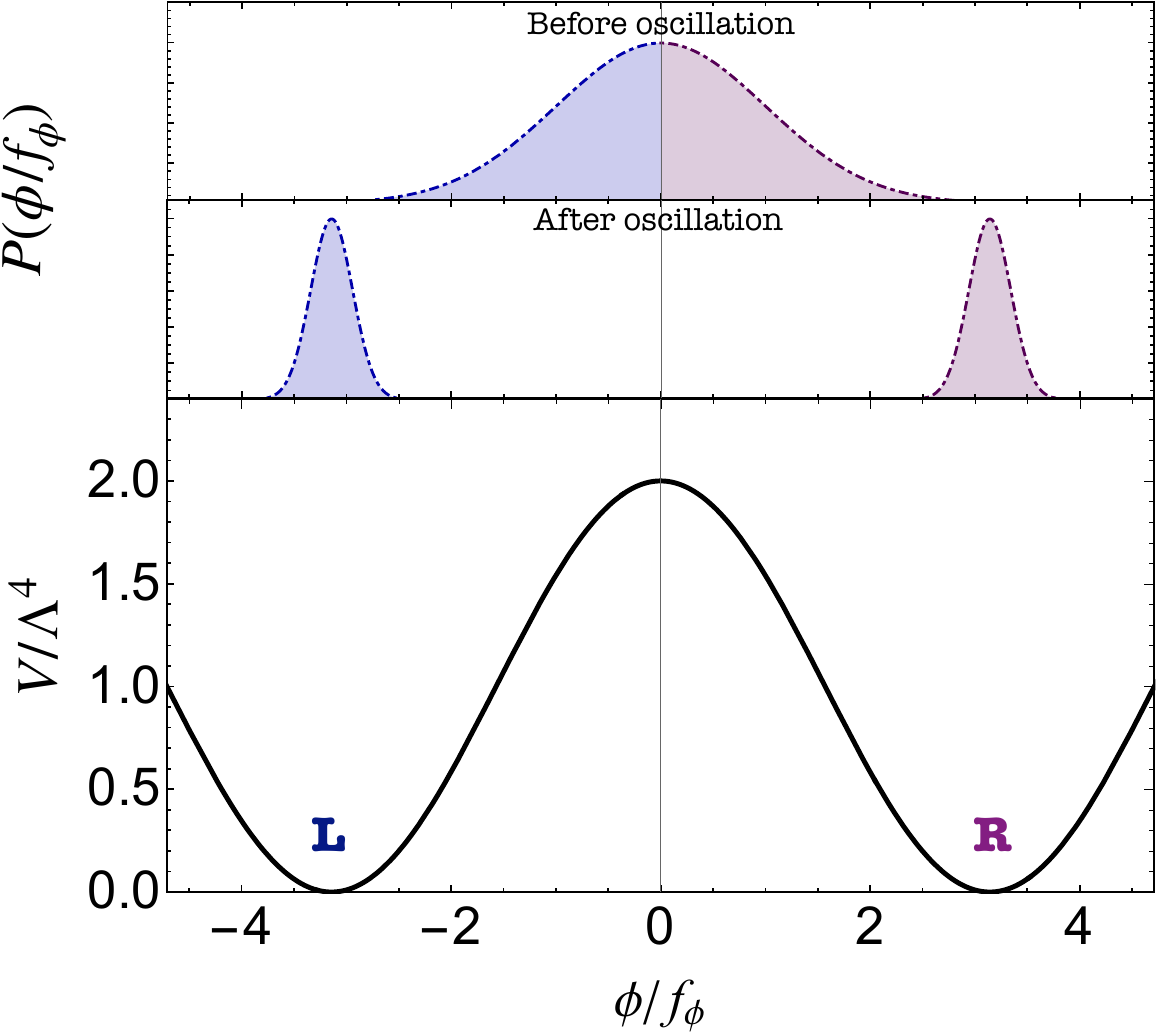}
      \end{center}
\caption{
The potential of \eqref{pot} is shown as a solid (black) line in the bottom panel.
The dot dashed line in the upper panel represents the probability distribution generated during the last $\O(10)$ e-fold of inflation. The axion in the blue (purple) region in the upper panel will settle down at the minimum $L$($R$) as in the middle panel, and these regions are separated by domain walls.
No strings are formed in this process. {Here we set  $\phi_0=0$ only for illustrative purpose, which is not assumed in the main text.} 
} \label{fig:1}
\end{figure}

For our purpose it is sufficient to consider the following potential,
\begin{eqnarray}\label{pot}
V(\phi) 
\,=\, \Lambda^4
\bigg[
1+\cos\bigg(\frac{\phi}{f_\phi}\bigg)
\bigg]
\,\simeq\, 2\L^4 -\frac{1}{2}  m_{\phi}^2  \phi^2+ \frac{1}{4} \left(\frac{\Lambda}{f_\phi}\right)^4 \phi^4 + \cdots
\end{eqnarray}
 where $\L$ is a dynamical scale,  $m_\f$ is the mass (curvature) of the axion, and the potential is expanded around the origin in the second equality.  See Fig.\,\ref{fig:1}.
 This potential is invariant under the discrete shift symmetry \eq{shift}, and as a result, there are degenerate vacua. 
 We focus on the two adjacent vacua at $\f=\pm \pi f_\f.$ 
Let us name the minimum $\f=-\pi f_\f$ as $L$ and $\f=+\pi f_\f$ as $R$.
Our argument does not depend on the precise shape of the potential, nor on whether the two minima are physically identical or not.

Let us assume that, during inflation, the axion is (almost) massless,\footnote{$\L$ may approach $0$ due to the finite temperature effect via Hawking radiation during inflation.
} i.e., $H_{\rm inf} \gg m_\phi$.
Then the axion acquires a quantum fluctuation, $\delta \phi = H_{\rm inf}/2\pi$, about a zero mode $\phi=\phi_0$, and it becomes classical after the horizon exit.
Here we define the zero mode by the axion field value averaged over the comoving scale corresponding to the Hubble horizon when the axion starts to oscillate much after inflation.
The fluctuations at superhorizon scales accumulate like a random walk, and the variance of the (gaussian) probability distribution of the axion field, $\sigma_\phi^2$, grows as $\sigma_\phi^2(N_e) = N_e( H_{\rm inf}/2\pi)^2$ with $N_e$ being the e-folding number. 
After inflation, the axion starts to oscillate when the Hubble parameter becomes around $H_{\rm osc} = m_\phi$. 

Suppose that $\phi_0$ is close enough to the origin and satisfies $|\phi_0| < \sigma_\phi(N_{\rm osc})$, 
where $N_{\rm osc}$ is the e-folding number when the fluctuation with the wavenumber $k/a = H_{\rm osc}$ exited the horizon during inflation. 
Then the probability distribution of the axion will be distributed across the potential maximum. According to the 
percolation theory~\cite{Vachaspati:1984dz,Vilenkin:1984ib}, if the probability of falling to the minimum $L$ (or $R$) is within the range of $0.31 \lesssim p \lesssim 0.69$,\footnote{When $p$ is out of this range,  many closed domain walls will be formed. They will soon collapse to evade the domain wall problem and may become black holes depending on the mass range~\cite{Khlopov:2004sc}. {In this case there will be no domain walls on the LSS if the axion starts to oscillate well before the reheating, and no KBCB is expected. Our mechanism in Sec.~\ref{sec:natural} can still be used to realize the initial conditions for such black hole formation.} } infinite domain walls are formed.
Thus,  domain walls without strings are likely formed if
\beq 
\laq{delphi}
\sigma_\theta \equiv \frac{\sigma_\phi(N_{\rm osc})}{f_\phi } = {\cal O}(1),
\eeq
or equivalently,
\beq
H_{\rm inf} \sim  f_\f,
\laq{maxim}\eeq  
where we have used $N_{\rm osc} = {\cal O}(10)$.
If the variance of the probability distribution is smaller by a factor of $\epsilon\, (<1)$, we would need an extra fine-tuning of order $\epsilon$ to set $\phi_0$ close enough to the origin. Here we assume the uniform prior probability distribution of $\phi_0$.\footnote{
Such a uniform distribution can be dynamically realized over very large scales if the duration of inflation is sufficiently long and the mass is sufficiently small. See also Refs.~\cite{Graham:2018jyp,Guth:2018hsa}
for the case where the axion mass plays an important role.
}  We will show later in this section 
that the condition \eq{delphi} is naturally satisfied in various UV models; e.g., with a non-minimal coupling to gravity,  $f_\phi$ is not a constant, but  dynamically set to a value close to $H_{\rm inf}$.\footnote{It is also possible to shift the axion potential by considering the inflaton-axion mixing~\cite{Daido:2017wwb,Takahashi:2019pqf, Takahashi:2019qmh, Nakagawa:2020eeg}. (See also \cite{Co:2018mho, Kobayashi:2019eyg,Huang:2020etx}). If the shift is equal or sufficiently close to $\pi$, we can effectively flip the sign of the potential. In this case, it is possible to set the axion distribution across the potential maximum if it is initially around the potential minimum.} Also, one can make use of fluctuations of another axion associated with the spontaneous symmetry breaking. In this case the condition for the formation of walls without strings is still given by \eq{delphi}, but the origin of fluctuation is different.

After inflation, the axion starts to oscillate when $H \sim H_{\rm osc} = m_\phi$. We assume that the axion starts to oscillate {before} the recombination, i.e.,
\beq
\laq{preCMB}
m_\f \; {\gtrsim} \; 3\times 10^{-29}\EV.
\eeq
We will discuss the case in which the axion starts to oscillate after the recombination in Sec.\ref{sec:dis}. Depending on the initial position,
the axion starts to roll down to either the minimum $L$ or $R$.
If the probability of falling to the minimum $L$ (or $R$) satisfies the above-mentioned condition, infinite domain walls are formed.  We emphasize that no strings are formed in this process, as the PQ symmetry is broken during inflation and never restored after inflation.
Such a symmetric phase does not exist in the case of string/M-theory axion. We will discuss the condition for the non-restoration of the PQ symmetry in the next subsection.

Infinite domain walls stretch over a Hubble radius due to its tension, while finite domain walls shrink and collapse once they enter the horizon. 
The resulting domain-wall network is known to follow the so-called  scaling solution for which there are on average $\O(1)$ domain walls in the Hubble horizon at any time. 
The scaling behavior of domain walls have been confirmed by numerical simulations~\cite{Press:1989yh,Garagounis:2002kt,Oliveira:2004he,Avelino:2005kn,Leite:2011sc,Leite:2012vn,Martins:2016ois}. Thus, the energy density of the domain wall scales as $\rho_{\rm DW} \sim \sigma_{\rm DW} H$
with $\sigma_{\rm DW}$ being the tension of the domain wall, and its typical curvature radius is of order the Hubble radius $1/H.$  {In the following, we assume that the scaling solution is valid until present, and we will come back to this issue in the last section.}

\begin{figure}[!t]
\begin{center}  
   \includegraphics[width=145mm]{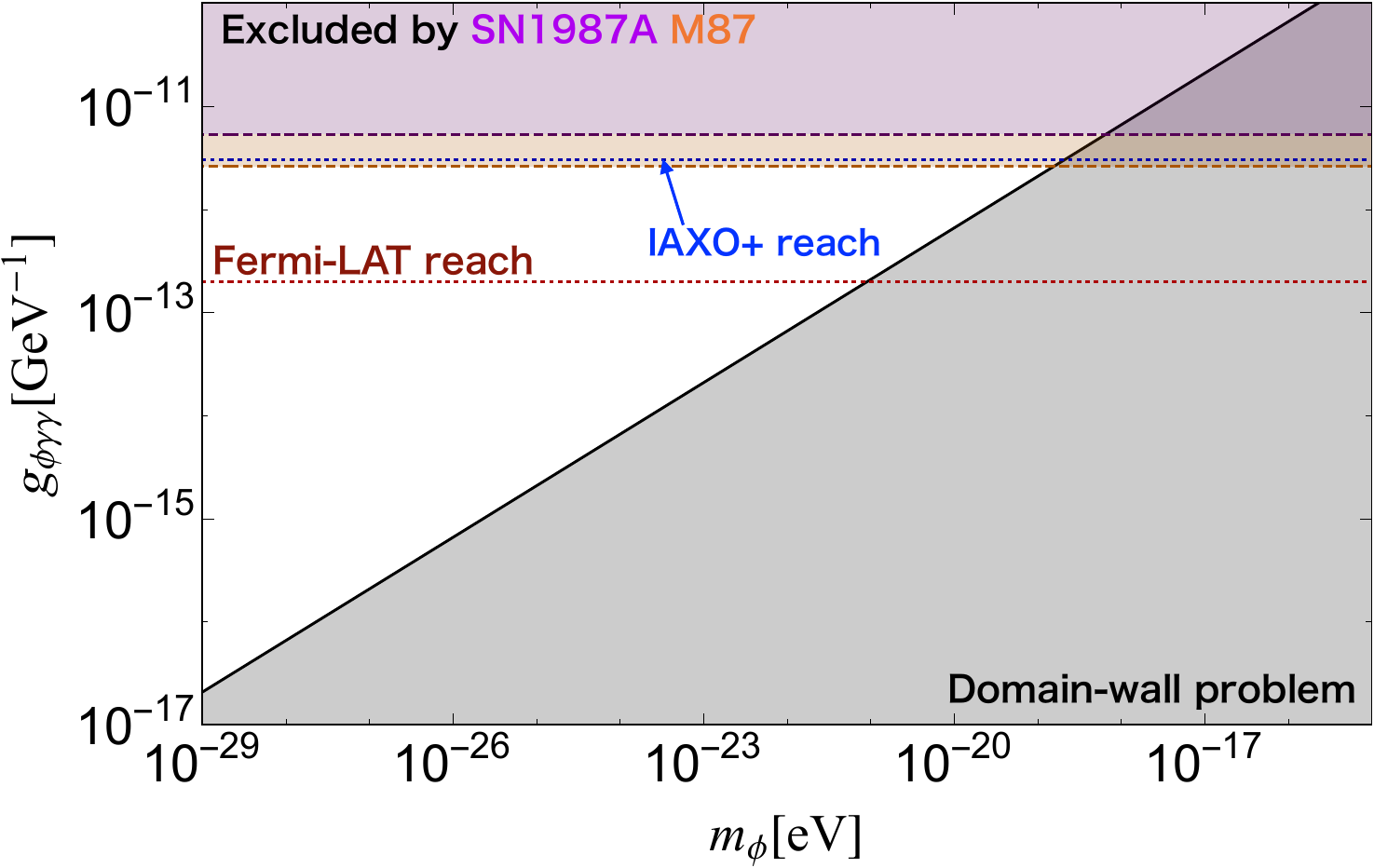}
      \end{center}
\caption{
Various constraints on the ALP domain walls without strings. The shaded regions are excluded.
The purple and orange regions above the two horizontal dashed lines are excluded due to the SN1987A (upper) and M87 (lower) bounds, respectively.
Also shown are the expected  IAXO$+$  and Fermi-LAT reaches in the blue (upper) and red (lower) dotted lines.
The lower right triangle (gray) region is excluded due to the domain wall problem where we have assumed $c_\g=1.$
We emphasize that the  predicted KBCB is universal in the entire white region. This should be contrasted to the scenario using a homogeneous ALP which requires a specific value of the ALP photon coupling for a given ALP mass. 
  } \label{fig:region}
\end{figure}
Since stable domain walls are formed before the recombination, we have the domain wall problem unless the tension satisfies~\cite{Zeldovich:1974uw,Vilenkin:1984ib},
\beq
\label{tension}
\sigma_{\rm DW} \simeq 8 f_\f^2 m_\f \lesssim (1 \MEV)^3,
\eeq
where the first equality is for the potential (\ref{pot}). 
This bound is set by the constraint from the CMB temperature fluctuation induced from the gravitational potential of the domain walls. 
This leads to 
\beq
f_\f \lesssim 4\times 10^{9}\GEV\sqrt{ \frac{10^{-20}\EV}{m_\phi}}. 
\eeq
Therefore, in our scenario we need to have a relatively small decay constant.
Various bounds on the parameter region are shown in Fig.\,\ref{fig:region}.
In the lower right triangle (gray) region, the aforementioned domain wall problem is serious. 
The bounds due to the lack of observation of the photon converted from the axion produced from SN1987A and the center of the radio galaxy M87 are  shown by the upper and lower dashed lines, respectively \cite{Payez:2014xsa,Marsh:2017yvc}. 
The region above the blue (upper) and red (lower) dotted line can be tested in the future by observing a Galactic core-collapse supernova with Fermi-LAT satellite \cite{Meyer:2016wrm} and  by observing solar axions via IAXO~\cite{Irastorza:2011gs,Armengaud:2014gea,Armengaud:2019uso}, respectively.
Note that, except for the IAXO reach and the domain wall problem bound, the shown constraints and sensitivity reaches depend on the assumed strength of the magnetic field in the Milky Way, and they may vary if the magnetic field is stronger or weaker than expected. 
The ALP-photon coupling may also be searched for by future measurements of CMB spectral distortions with certain primordial magnetic fields~\cite{Tashiro:2013yea}. As we shall see in the next section, both isotropic and anisotropic CB are predicted in the white region.

\subsection{Validity of the EFT description, and  UV completion}
\label{sec:cons}

So far we have used the ALP potential (\ref{pot}) as an effective theory (EFT) to describe our mechanism. 
Our mechanism can have several UV completions such as the string/M-theory 
or a renormalizable field theoretic axion model. 
One of the important requirements for our mechanism is that the PQ symmetry should not be restored, since otherwise cosmic strings are produced, which induce only anisotropic CB~\cite{Agrawal:2019lkr}.
This actually depends on the UV completion, and here we study if it is satisfied in explicit UV models.

First of all, let us study the condition that we can describe the whole thermal history within the EFT with the higher dimensional term of \eq{int}. 
The validity of the EFT requires that the maximum photon temperature, $T_{\rm max}^{\rm th}$, after inflation should satisfy
\beq
\laq{ineq 1}
T_{\rm max}^{\rm th}< T_{\rm max}^{\rm EFT}\sim g^{-1}_{\f\g\g} = 4\times 10^{11}\GEV\( \frac{ f_\f/c_\g}{10^{9}\GEV}\). 
\eeq
where  $T_{\rm max}^{\rm EFT}$ is set so that the axion photon interaction is perturbative. 
In the thermal history of the universe, as in the case of simple exponential decay,
the temperature at the beginning of reheating is likely to be the highest.
The maximum temperature in this case is given by
\beq
T_{\rm max}^{\rm th}\sim  \(\frac{\pi^2 g_\star}{30}\)^{-1/4} \(\rho_{\f} \times \frac{\G_{\f}}{H_{\rm inf}}\)^{1/4} 
\eeq 
where $\Gamma_\phi$ is the (constant) decay rate, 
$g_\star$ is the relativistic degrees of freedom in the thermal plasma, and
the inflaton energy density at the beginning of the reheating is $\rho_\f \sim 3({M_{\rm pl} H_{\rm inf}})^2$.
The inequality \eq{ineq 1} can be satisfied if 
\beq
\laq{consistency}
\G_{\f}\lesssim 50\GEV c_\g^{-4} \(\frac{f_\f}{10^9\GEV} \)^4 \(\frac{10^9\GEV}{H_{\rm inf}}\).
\eeq
If $\G_\f$ is a constant in time until the completion of the reheating, we obtain the reheating temperature $T_R= 6\times 10^9\GEV(\frac{\Gamma_\f}{50\GEV})^{1/2} .$
 The ALPs can be produced due to thermal scattering via the coupling to photons. As a result, thermally produced ALPs contribute to the effective neutrino number of $\D N_{\rm eff}$ (see e.g. Ref.~\cite{Salvio:2013iaa}). If $T_R\sim f_\f$, the ALPs are completely thermalized, we have $\D N_{\rm eff}\sim 0.03$. Such thermalized ALP can be searched for in the future CMB and baryonic acoustic oscillation experiments~\cite{Kogut:2011xw, Abazajian:2016yjj,Baumann:2017lmt}.

{A UV theory} for the ALP is required for {a perturbative description} if  \eq{consistency} is not satisfied. As a {simple UV model},
let us consider a model with a $\U(1)_{\rm PQ}$ symmetry by introducing
 a complex PQ scalar $S$ with the potential 
\beq V_S=-m_{S}^2 |S|^2 + \lambda |S|^4/2,\eeq 
where $m_S$ is the mass parameter, and $\l$ is a quartic coupling constant. The potential preserves the phase rotation of $S$, which is identified with the U(1)$_{\rm PQ}$ symmetry.
If $m_S^2>0$, $S$ acquires a vacuum expectation value (VEV)  $\vev{S}=
f_\f/\sqrt{2} \equiv
m_S/\sqrt{\lambda}$, and 
the Nambu-Goldstone boson (NGB) of the spontaneous broken $\U(1)_{\rm PQ}$ is the axion, $\f$, which resides in the phase of $S$. The axion
 mass may be generated by a tiny explicit breaking term of the PQ symmetry, like $\d V_S= \k S+{\rm h.c.}=\k \sqrt{2}f_\f \cos{(\f/f_\f+\arg{\k})}$, where $\kappa$ parametrizes the small breaking.\footnote{
 If we introduce a breaking term $\propto S^n$, the period of the potential rather than the VEV of $S$ should be regarded as the definition of $f_\phi$.
 } One can clearly see that the axion potential in this case respects the discrete shift symmetry \eq{shift}. Although there is a unique vacuum in this example, the physically identical vacua separated by the domain wall can be distinguished in the absence of strings, and such domain walls are stable.\footnote{Except for quantum creation of a string loop on the wall, which is however exponentially suppressed.}
 
To obtain the axion coupling to photons, let us introduce a pair of PQ  chiral fermions, $\p$, carrying the electromagnetic charge $q$ (given by a combination of $\U(1)_{\rm Y}$ and $\SU(2)_{\rm L}$ charges) with the following interaction:
\beq
{\cal L}^{\rm UV}\supset y S \bar{\p}\hat{P}_L\p  + {\rm h.c.},
\eeq
where we denote the chiral projection operator by $\hat{P}_{L(R)}$.
We notice that $\p$ should not induce the $\U(1)_{\rm PQ}\SU(3)^2_c$  anomaly which would generate a much heavier mass for $\f$ due to nonperturbative QCD effects a la the QCD axion.\footnote{We can relax this condition by introducing the QCD axion separately~\cite{Anselm:1981aw,Pospelov:2008gg}. See also discussion in Sec.~\ref{sec:qcdaxion}} 
Here we assign a unit PQ charge on $S$  and $\hat{P}_R \psi$.
Through this interaction the fermion $\psi$ gets a mass of \beq 
m_\p=y \frac{f_\f}{\sqrt{2}}\sim 700\GEV\( \frac{y}{10^{-6}} \)\(\frac{f_\f}{10^9\GEV}\)
\eeq
due to the VEV of $S$.

Since there is a $\U(1)_{\rm PQ} \U(1)_{\rm em}^2$ anomaly, we obtain $c_\gamma= q^2 N_\p$ by integrating out $\p$ where $N_\p$ is the multiplicity of $\p.$
The fermion is produced due to thermal scattering during the reheating via the gauge interaction. 
Furthermore, the thermal corrections from fermion loops give a positive mass squared to the PQ field 
\beq
\d m_{S}^2\sim y^2 T^2.
\eeq
For the non-restoration of the PQ symmetry, we need
\beq
\d m_{S}^2\lesssim f_\f^2 \to T_{\rm max}^{\rm th }\lesssim 10^{15}\GEV \(\frac{10^{-6}}{y}\)\(\frac{f_\f}{10^9\GEV}\).
\eeq
where we have assumed the vacuum mass of the PQ Higgs to be around $m_S\sim f_\f$ (i.e. $\lambda \sim 1$). 

In this model there is unbroken $Z_2$ parity under which the fermion $\psi$ flips the sign. Thus, unless we introduce other interactions, the fermion $\psi$ is stable.
Such stable particle may cause a cosmological problem if it is too abundant. However if the mass is around TeV range, we can have the right relic abundance for dark matter by the WIMP mechanism. 
In fact, depending on the electroweak charge assignment,
the lightest neutral component may become the dominant dark matter like the Wino or Higgsino.

At high temperatures the axion-photon-photon coupling is suppressed as 
$\sim \a |m_\p|^2 /(f_\f T^2)$ by the typical energy$\sim T$ of $\p$ running  in the triangle diagram. 
Still the thermal axion may be produced through the coupling with $\psi$ and contribute to $\D N_{\rm eff}$~(see e.g. \cite{Daido:2017wwb, Daido:2017tbr, Takahashi:2019qmh, Takahashi:2020uio}). 

\subsection{A model with a negative Hubble mass term}
\label{sec:natural}
So far we have seen that domain walls without strings 
are likely formed if the ALP fluctuation is comparable to its periodicity of the potential, i.e. if \eq{delphi} or \eq{maxim} is satisfied. 
The question is whether this condition is satisfied naturally. 
Here we show that, by using the UV completion with the PQ scalar in the previous subsection,\footnote{Also, the field theoretic axion can be realized in string theory. } the condition \eq{maxim} can be naturally satisfied.
The key ingredient is 
 the non-minimal coupling to gravity,
\beq
{\cal L} \supset\sqrt{-g}  R \left(\frac{\xi}{2}|S|^2+\frac{M_{\rm pl}^2}{2}\right)
\eeq
where  $g$ is the determinant of the metric and $\x (> 0)$ is the non-minimal coupling.
During  inflation, this coupling gives an effective mass term to $S$ as\footnote{We may also consider a thermal correction from the Gibbons-Hawking radiation. For instance, the thermal mass squared, $\sim \lambda H_{\rm inf}^2/(2\pi)^2,$ via the quartic term may be generated. However this is smaller than the Hubble-induced mass term considered in the text if $\l\sim \x$. } 
\beq
-6 \xi H_{\rm inf}^2 |S|^2,
\eeq
which is often called the Hubble-induced mass term.
If this dominates over the bare mass term of $m_S^2$, the PQ field obtains an expectation value during inflation as
\beq
\vev{S}_{\rm inf} \simeq \sqrt{\frac{6 \xi}{\lambda}}  H_{\rm inf}. 
\eeq
The decay constant of the axion in this period, which we denote $f_{\f,{\rm inf}}$, is different from that in the vacuum, $f_\f$. Here
\beq
f_{\f,{\rm inf}} \simeq 2\sqrt{\frac{3\xi}{\lambda}} H_{\rm inf}. 
\eeq
Then we get a variance during inflation on the misalignment angle given by
\beq
\sigma_\theta^2
=
\frac{\sigma_\phi^2}{f_{\f,{\rm inf}}^2} \sim \frac{\xi}{\lambda},
\eeq
which is of $\O(1)$ if $\xi \sim \lambda$.
The most natural values of $\xi$ and $\lambda$ are of order unity, and so, 
the fluctuation of the misalignment angle is order one in this case.

After the inflation, the PQ scalar still receives the negative Hubble-induced mass term when the universe is dominated by the inflaton coherent oscillations. Thus, the VEV of $S$ decreases as $H$, and at a certain point, it will become equal to $f_\phi/\sqrt{2}$ when the bare mass term dominates over the Hubble-induced mass term.
Note that
the Hubble-induced mass term gets suppressed in the radiation dominated era due to the (approximate) conformal invariance. Thus, the VEV of $S$ should settle down at $f_\phi/\sqrt{2}$ before the reheating completes. 
This is the case 
if \eq{consistency} is satisfied. Note also that,
as long as the PQ symmetry is never restored after inflation, the variance of the misalignment angle distribution {at large scales} is preserved until the onset of the oscillations of the ALP. {At small scales, on the other hand, the axion fluctuation gets enhanced as the effective decay constant decreases, which will also help to form domain walls without strings~\cite{Kobayashi:2016qld}.
As a consequence,  the condition \eq{delphi} or \eq{maxim} is naturally satisfied in this set-up, and domain walls without strings can be formed.

\subsection{A model with the QCD axion and ALP}
\label{sec:qcdaxion}
Here we consider an alternative mechanism to populate the ALP field over a certain range to form domain walls without strings. 
To this end we introduce two axions with a mixing. 
The idea is to transfer  fluctuations of one of the axions due to spontaneous symmetry breaking, to the ALP by using the mixing effect.
To be concrete, the heavy and light mass eigenstates are identified with the QCD axion and ALP, respectively. Then, axionic strings due to the spontaneous PQ breaking for the heavier axion disappear around the QCD phase transition, but some part of the fluctuations remain in the ALP.

For simplicity let us consider a model with two PQ scalar fields, $\Phi_1 \AND \Phi_2$, which are charged under global $\U(1)_{\rm PQ,1}$ and $\U(1)_{\rm PQ,2}$ symmetries, respectively. In the vacuum both $\Phi_1$ and $\Phi_2$ are assumed to develop a nonzero VEV, leading to two NGBs, $\varphi_1$ and $\varphi_2$.  We introduce the following interactions,
\beq 
{\cal L}\supset y_1 \varphi_1 \bar{Q}_1 \hat{P}_L Q_1 + y_2  \varphi_2 \bar{Q}_2 \hat{P}_L Q_2+{\rm h.c.},
\eeq
where $Q_1$ and $Q_2$ are the PQ quarks with certain PQ charges, and we assume that they are in the fundamental representations of SU(3)$_C$.  If the PQ quarks are also charged under $\SU(2)_L \times \U(1)_Y$, the ALP will have a coupling to photons. Here and in what follows we suppress the multiplicity of the PQ quarks for a concise notation.

In the low energy, both $Q_1$ and $Q_2$ acquire a heavy mass due to the VEVs of 
$\Phi_1$ and $\Phi_2$. Integrating out the PQ quarks, we obtain couplings of $\varphi_1$ and $\varphi_2$ to gluons through the QCD anomaly,
\begin{align}
    \frac{g_s^2}{32 \pi^2}
    \left(\frac{\varphi_1}{f_1} + \frac{\varphi_2}{f_2}\right)
    G_{a \mu \nu} \tilde{G}_a^{\mu \nu},
    \label{phigg}
\end{align}
where $g_s$ is the strong gauge coupling, the decay constants $f_1$ and $f_2$ are defined through the above equation; in other words, the multiplicity of the PQ quarks is included in the definition of the decay constants. If the two global U(1) symmetries are explicitly broken only by the QCD anomaly (\ref{phigg}), 
the combination shown in the parenthesis is identified with the QCD axion, $a$,
\beq
\frac{a}{f_a}= \frac{\varphi_1}{f_1}+\frac{\varphi_2}{f_2}
\eeq
with $f_a^{-2}\equiv{f_1^{-2}+f_2^{-2}}$, while the  orthogonal one remains massless and it is identified with the ALP $\phi$,
\begin{align}
    \frac{\phi}{f_a} = \frac{\varphi_1}{f_2}-\frac{\varphi_2}{f_1}.
\end{align}
By turning on another tiny symmetry breaking, we can give a very small mass to $\phi$, and it remains an approximately light mass eigenstate as long as it is much lighter than the QCD axion. Note that the
 periodicity of the potential along $\phi$ determines the decay constant $f_\phi$. For instance, we may couple $\Phi_2$ to hidden quarks charged under hidden QCD which becomes strong at low energy.\footnote{{This is particularly the case if the PQ symmetries are accidental ``baryon number" symmetries of chiral hidden QCD~\cite{Lee:2018yak}. See also Refs.\,\cite{Randall:1992ut, DiLuzio:2017tjx, Ardu:2020qmo, Yin:2020dfn}. In the case of $\SU(N_i)$ gauge symmetry with chiral fermions, the resulting PQ symmetry is generically anomalous to $\SU(N_i)$. If the PQ symmetry and $\SU(N_i)$ is spontaneously broken at the same time, which happens in many cases, e.g. the PQ Higgs 
 charged under $\SU(N_i)$ is a symmetric bi-fundamental tensor, a small instanton will generate a tiny mass to the ALP/axion. 

 }
} Then $\varphi_2$ acquires an extra potential due to  non-perturbative effects of the 
 hidden QCD. If the adjacent minima of the extra potential  are related by $\varphi_2 \to \varphi_2 + 2 \pi p f_2 $ with $p$ being a rational number, the decay constant of $\phi$ is given by
 \begin{align}
     f_\phi = \frac{f_1}{p f_2} f_a
 \end{align}
so that the adjacent vacua along $\phi$ are related by
$\phi \to \phi + 2 \pi f_\phi$.

Let us assume that 
$\U(1)_{\rm PQ,1}$ is restored during inflation
and gets spontaneously broken after inflation, 
while $\U(1)_{\rm PQ,2}$ is already broken during inflation and never restored afterwards. We also assume that
the domain wall number of  $\varphi_1$ satisfies $N_{\rm DW}^{(1)}=1$, while there is no constraint on the domain wall number of $\varphi_2$, $N_{\rm DW }^{(2)}$. In other words, $\varphi_1$
and $\varphi_1+2\pi  /f_1$ are assumed to be physically identical.\footnote{In a more general setting where there are multiple NGBs that make up the QCD axion, one of their domain wall numbers must be equal to one.}
After the spontaneous breaking of $\U(1)_{\rm PQ,1}$, 
 $\varphi_1$ randomly takes values between $-\pi$ and $\pi$ in each Hubble horizon (or a domain with the size of the correlation length). Then, 
there appear cosmic strings of $\Phi_1$, which will soon follow the scaling solution. On the other hand, $\varphi_2$ takes a fixed value $\varphi_2 = \varphi_{2i}$ except for a quantum fluctuation around it. 

Around the QCD phase transition, the non-perturbative QCD effect  generates a potential for the QCD axion $a$. Noting that the QCD axion changes by $2\pi f_a$ around the cosmic string of $\Phi_1$,  it can be regarded as the axionic string. Since the domain wall number $N_{\rm DW}^{(1)}$ is equal to unity, a single domain wall is attached to each cosmic string, and the string-wall network  soon disappear after the QCD axion starts to oscillate. 
The right amount of QCD axion to explain dark matter can be
produced from this process for $f_a = {\cal O}(10^{11})$\,GeV according to the recent numerical simulations~\cite{Fleury:2015aca,Klaer:2017ond,Gorghetto:2018myk,Vaquero:2018tib,Buschmann:2019icd,Gorghetto:2020qws,Kawasaki:2018bzv}.

We can treat the ALP as a massless axion during the QCD phase transition, as the ALP mass of our interest is much lighter than the QCD axion. Interestingly, the ALP $\phi$ inherits a part of fluctuations from $\varphi_1$, and it fluctuates
around $\phi = - \frac{f_a}{f_1} \varphi_{2i}$ with the width of $\pm \pi f_1^2/\sqrt{f_1^2+f_2^2}$. Thus, if there is no large hierarchy in the decay constants, the ALP acquires a sizable fluctuation comparable to its decay constant, leading to the formation of domain walls without strings.


Lastly let us discuss a case in which both 
 $\U(1)_{\rm PQ,1} \AND \U(1)_{\rm PQ,2}$ get
 spontaneously broken after inflation. 
After the symmetry breaking, there appear two kinds of cosmic strings for $\Phi_1$ and $\Phi_2$.
Each type of cosmic strings will soon follow the scaling solution. 
Around the QCD phase transition, the QCD axion acquires a potential, and there appear domain walls attached to these strings.
While a single domain wall is attached to the string of $\Phi_1$ because of $N_{\rm DW}^{(1)}=1$, $N_{\rm DW}^{(2)}$ domain walls will be attached to the string of $\Phi_2$. Due to the tension of domain walls, some of the strings and walls disappear, and we are left with bundles of strings composed of $N_{\rm DW}^{(2)}$ strings of $\Phi_1$ attached to a string of $\Phi_2$. Such a bundle of strings corresponds to the 
cosmic strings for the ALP $\phi$, and it was studied in details in a context of the clockwork/aligned QCD axion model~\cite{Higaki:2016jjh,Long:2018nsl}.
Thus, in this case, its prediction for the CB will be similar to the scenario of Ref.~\cite{Agrawal:2019lkr}.

When $\U(1)_{\rm PQ,1}$ remains broken but $\U(1)_{\rm PQ,2}$ is restored, on the other hand, there is a cosmological domain wall problem associated with the  cosmic strings for $\Phi_2$ and $N_{\rm DW}^{(2)}$ domain walls attached to them, unless $N_{\rm DW}^{(2)} = 1$.
If both $\U(1)_{\rm PQ,1}$ and $\U(1)_{\rm PQ,2}$ remain broken during and after inflation, we do not have cosmic strings and there is no domain wall problem due to the QCD phase transition. 
The domain wall without strings of the ALP may be generated due to the inflationary fluctuation 
as we have discussed in the previous subsection.
In this case, however, there may be an isocurvature and domain wall (without strings) problems for the QCD axion. The problem will be relaxed if $f_\phi$ is hierarchically smaller than $f_a$.

To sum up, we have found that it is possible to 
populate a light ALP over a certain field range in a scenario using two NGBs in which one of the global U(1) symmetries gets spontaneously broken {after inflation and the corresponding cosmic strings disappear due to the explicit breaking of a combination of the U(1) symmetries}.
{Then,} even after the strings and walls disappear due to their tension, there remain the field fluctuations along the light ALP due to the mixing of the two NGBs. Note that there is no sizable isocurvature perturbation in this scenario.
As in the scenario with a negative Hubble mass, it is possible to  generate ALP domain walls without strings if there is no large hierarchy in the parameters.

\section{Kilobyte Cosmic Birefringence}
\label{sec:KB}
Now we come to our main point. In the presence of the domain walls today, we are either in the vacuum $L$ or $R$. 
Suppose we are in the vacuum $R$.
The photon emitted  from the LSS in the direction of $\Omega$ reaches us by going through many domain walls along the line of sight. Since the width of each domain wall 
$\sim 1/m_\f$ is much larger than the typical wavelength of the CMB photon (See Fig.\,\ref{fig:region}), the polarization plane rotated adiabatically following \Eq{delF}. 
Therefore, the polarization angle changes  by 
$\pm c_\g \a$ each time the photon passes through the wall.

In principle, we can estimate $\D\F(\Omega)$ by summing over all the contributions from the domain walls that the CMB photon passes through. However, the net
 $\D\F(\Omega)$ is simply determined by the difference of the ALP field value between the domain on the LSS and us, and the 
 detailed information of walls on the way is absolutely irrelevant, because their contributions are  canceled out.
 The total shift of the angle, $\D \F(\Omega)$, depends only  on in which vacuum the last scattered photon (LS$\g$) was emitted. In other words, we have
\begin{align}
\D\F &=0 \text{      ~~~~~       if  LS$\g$ is from the vacuum $R$}\non \\\laq{polar}
\D\F &= c_\g \a \text{   ~~~if  LS$\g$ is from the vacuum $L$}.
\end{align}
Note that the volume (and therefore the area on the LSS) occupied by the $L$ and $R$ vacua are almost equal since they are equivalent. The symmetry is spontaneously broken by the choice of the vacuum at the location of the solar system, and we have chosen our vacuum to be $R$. Thus,
by averaging the two cases, we get the isotropic CB,
\beq
\beta = \frac{1}{4 \pi}\int d\Omega\, \Delta \Phi(\Omega) = \frac{1}{2} c_\gamma \alpha  \simeq 0.21 c_\g\,  {\rm deg}.
\eeq
Intriguingly, it is consistent with the recently reported value of \eq{measure} if $c_\g=\O(1)$.

According to the scaling solution of the domain-wall network, each Hubble patch should contain $\O(1)$ domain walls on average. 
Since the LSS contains 
$\O(10^3)$ Hubble patches, we expect a similar number of domain walls separating the two vacua $L$ and $R$.
The adjacent vacua separated by the domain wall has different $\D\F$ in \Eq{polar}. 
Consequently, each domain on the LSS has  1-bit information, i.e., $\Delta \Phi = 0$ or $c_\g \a$.
Since there will be  $\O(10^{3-4}) $ domain walls at the LSS, we call it as the kilobyte cosmic birefringence (KBCB).

In addition to the isotropic CB, anisotropic one with a peculiar feature is also generated. In particular, such anisotropic CB directly reflects the domain-wall configuration at the LSS; it is expected to peak at scales corresponding to the typical size of domains ($\sim$ the Hubble horizon), while it is suppressed at larger and smaller scales. The reason why it is suppressed at larger scales is due to 
 the scaling solution; each Hubble horizon looks similar on average. The reason why it is suppressed at smaller scales is the scaling nature of
 domain walls.

 The typical magnitude of the anisotropic CB on the scales at the peak is expected to be slightly smaller than the isotropic CB. This can be seen as follows. First let us define the anisotropic part of the net rotation of the polarization angle,
 \begin{align}
     \Delta \tilde{\Phi} \equiv 
     \Delta \Phi - \beta.
 \end{align}
 Since $\Delta \tilde{\Phi}$ takes a value of $\pm c_\gamma \alpha/2$, we obtain
 \begin{align}
 \label{phi2}
      \int{d\Omega \left(\Delta \tilde{\Phi}(\Omega)\right)^2}\simeq 4\pi \beta^2,
 \end{align}
where we have used the fact that the vacua L and R should occupy almost the same area on the LSS. One can expand $\Delta \tilde{\Phi}$ in terms of the spherical harmonics,
\begin{align}
    \Delta \tilde{\Phi}(\Omega) = \sum_{\ell,m}
    a_{\ell m} Y_{\ell m}(\Omega),
\end{align}
{where the expansion coefficients satisfy $a_{\ell m}^*=a_{\ell -m}$ 
because $\Delta \tilde{\Phi}(\Omega)$
is a real parameter. The above expansion can be inverted as follows,
\begin{align}
\label{alm}
    a_{\ell m} = \int d\Omega \,  \Delta \tilde{\Phi}(\Omega) Y_{\ell m}^*(\Omega).
\end{align}
}
We define the angular power spectrum as
\begin{align}
\label{aps}
    C_\ell^\Phi \equiv \frac{1}{2\ell + 1} \sum_m 
    a_{\ell m}^* a_{\ell m} 
\end{align}
{without taking an ensemble average.
We will distinguish the angular power spectrum with the ensemble average
by adding a bar.}
Then we can express the lhs of
(\ref{phi2}) as
\begin{align}
  \int{d\Omega \left(\Delta \tilde{\Phi}(\Omega)\right)^2}
  =
    \sum_{\ell}{(2\ell+1) C^\F_\ell}.
\end{align}
{Before we estimate the angular power spectrum based on a model of the domain-wall network, let us make an order estimate of its upper limit.
Assuming that} the angular power spectrum has a relatively broad peak around $\ell = \ell_p$
with a width
$\delta \ell_p \sim \ell_p$,
we obtain 
\begin{align}
\laq{peak}
    \frac{ \ell_p(\ell_p+1)C^\F_{\ell_p}}{ 2\pi}\lesssim 0.1\, \(\frac{\beta}{{0.35} \,{\rm deg}}\)^2 {\rm deg}^2.
\end{align}
{
As we shall see shortly,  the predicted power spectrum is significantly deviated from the scale-invariant one, and so, 
the above estimate cannot be directly compared to the upper bound on a scale-invariant anisotropic CB, $\ell(\ell+1)C_\ell /2\pi < 0.033{\rm\,deg}^2$ (95 \% CL)~\cite{Bianchini:2020osu}. In fact, we will see that the predicted anisotropic CB is consistent with the current observations.}

{For a more precise estimate of the angular power spectrum, let us model the domain wall network at the recombination as follows.
We neglect the thickness of the domain wall for simplicity, which is a good approximation well after the domain walls has been formed. Now, if we consider a straight line along an arbitrary direction at recombination, there will be many domain walls along the line. 
Let us denote the density of domain walls along the line by $P_{\rm DW} \equiv r_{\rm DW}^{-1}$, where
$r_{\rm DW}$ is the average distance between the adjacent domain walls. For a scaling solution, it is considered to be given by 
\begin{align}
    P_{\rm DW} = \kappa_{\rm DW}  H,
\end{align}
with $\kappa_{\rm DW}$ being a numerical coefficient of $\O(1)$. 
The precise value of $\kappa_{\rm DW}$ can be determined by a dedicated numerical simulation, but here we treat it as a free parameter of $\O(1)$.
}

{Let us choose a sufficiently short interval $\delta x$ so that
 the probability that a domain wall exists in the interval is given by $\delta x  P_{\rm DW} \ll 1$.
We can then estimate a probability that after traveling a distance $\Delta x=N\d x$ the vacuum remains the same by the Poisson distribution,
\beq
P_{\rm match}[\Delta x]=\sum_{N/2\geq m=0} \frac{N!}{2m! (N-2m)!}(P_{\rm DW} \delta x)^{2m} (1-P_{\rm DW} \delta x)^{N-2m}\to \frac{1}{2}(1+e^{-2P_{\rm DW} \Delta x}),
\eeq}
where $2m$ denotes the number of domain walls in the interval $\Delta x$, and we take a limit of $N \to \infty$ and $\delta x \to 0$ for  fixed $\Delta x$.}
By using this formula, we can calculate the ensemble average of the two-point function as
\beq
\vev{\Delta \tilde{\F}(0,0) \Delta \tilde{\F}(\theta, 0)}\simeq \beta^2  (2 P_{\rm match}-1) = \beta^2  e^{-2  P_{\rm DW} R \sqrt{2(1-\cos\theta)}},
\label{tpt}
\eeq 
{where we explicit show the polar coordinates $\Omega = (\theta,\varphi)$, and assume the statistical isotropy.
Here $R\sqrt{2(1-\cos\theta)}$ in the exponent represents the physical distance between the two points at the time of last scattering. For the standard cosmological parameters~\cite{Aghanim:2018eyx}, the exponent of (\ref{tpt}) 
is approximately given by $\sim 176 \kappa_{\rm DW} \sqrt{1-\cos\theta}$.
}
{Using (\ref{alm}) and (\ref{aps}) with an ensemble average, we obtain }
\beq 
\ol{C}_{\ell}^\F=2\pi \int{d \cos\h \vev{\Delta\tilde{\F}(0,0)\Delta \tilde{\F}(\theta,0)}P_\ell(\cos\h)},
\label{apsen}
\eeq 
with $P_\ell$ being the Legendre polynomial. 

\begin{figure}[!t]
\begin{center}  
   \includegraphics[width=145mm]{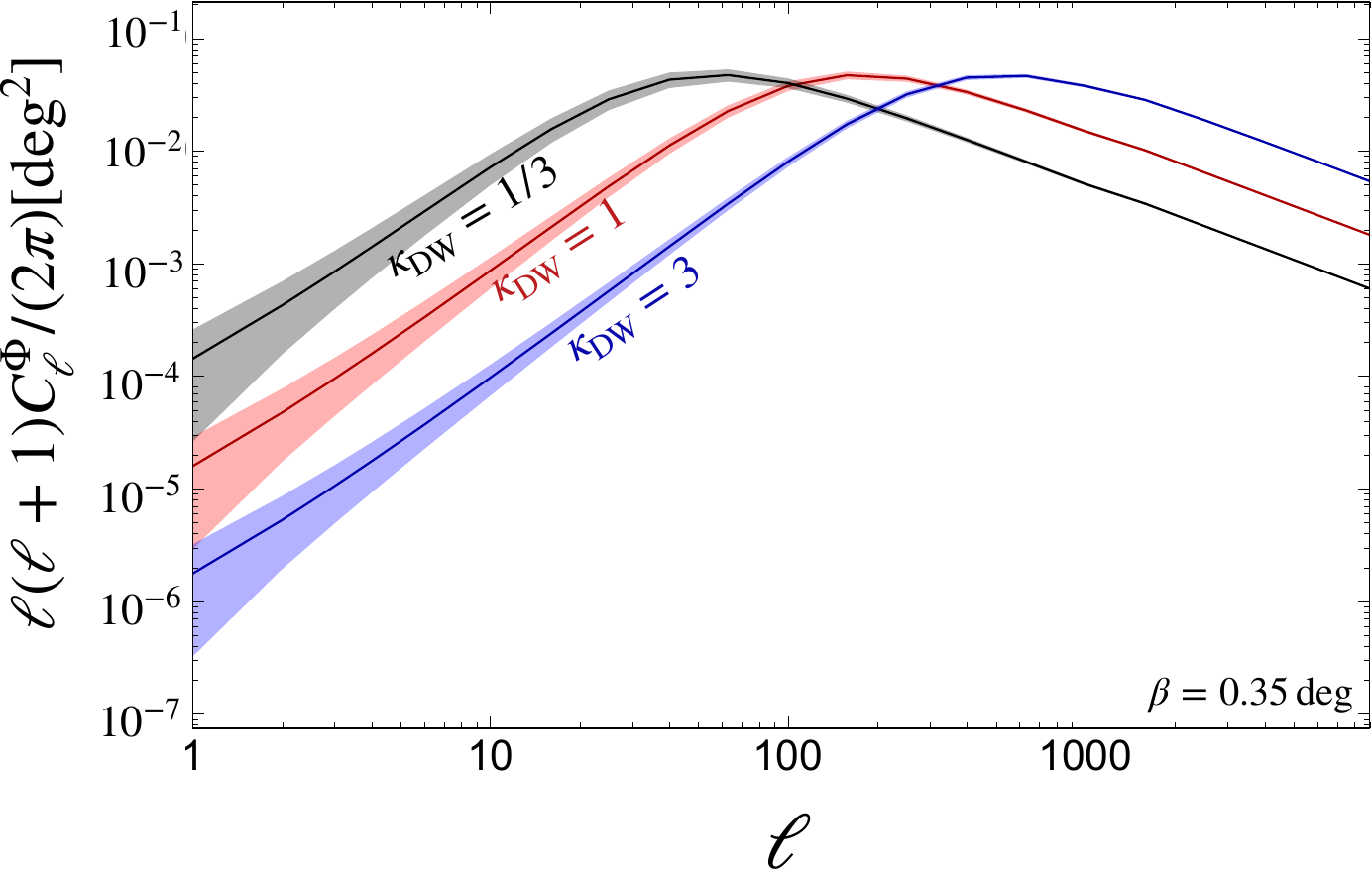}
      \end{center}
\caption{{The angular power spectrum of the anisotropic CB in the KBCB scenario. Here we set $\beta=0.35$deg, and vary $\k_{\rm DW}=1/3,1$, and $3$ from left to right. The bands represent the cosmic variance  $\Delta C^\F_\ell\approx \sqrt{\frac{2}{2\ell +1}} \ol{{C}}^\F_\ell $ (see Appendix for derivation).}} \label{fig:Cl0}
\end{figure}

{By substituting (\ref{tpt}) into (\ref{apsen}), we can estimate the anisotropic CB predicted in our KBCB scenario. The results are shown in Fig.~\ref{fig:Cl0}, where we use the large $\ell$ expansion of $P_\ell(\cos\theta)$ for $\ell>1000$ to reduce the numerical cost.
In Fig.~\ref{fig:Cl0}, we fix the isotropic CB as $\beta=0.35$deg, and set {$\k_{\rm DW}=1/3,1,\AND 3$}
from left to right.
The bands represent the cosmic variance, $\Delta C^\F_\ell \approx \sqrt{\frac{2}{2\ell+1}} \ol{C}^{\F}_\ell$, which is considered to be a good approximation as long as $\ell$ is not much larger than  $P_{\rm DW}R$ (see Appendix).
 As expected, the angular power spectrum of the anisotropic CB, $\ell(\ell+1) \ol{C}^\F_\ell/(2\pi)$, has a characteristic peak at $\ell_p \sim P_{\rm DW} R$. Interestingly, the peak height is almost  independent of $\k_{\rm DW}$. Numerically we find 
\beq 
\frac{\ell_p(\ell_p+1) \ol{C}^\F_{\ell_p}}{2\pi}\sim 0.05\, {\rm deg^2}  \(\frac{\beta}{0.35{\rm deg}}\)^2,
\label{apsp}
\eeq
which saturates the rough estimate of \eq{peak}. The dependence of the peak height on $\kappa$ can be seen as follows. The integration of (\ref{apsen}) receives dominant contributions from
 $0 \leq \theta \lesssim (P_{\rm DW} R)^{-1} \ll 1$ due to the exponential factor of the two-point correlation function. 
 {At small $\theta, \ell \theta  \ll 1$, the Legendre polynomial can be approximated by $P_\ell(\cos\theta) = 1 - \ell (  \ell+1)\theta^2/4+\O(\ell^4 \theta^4).$}
Similarly, {at small $\theta \ll 1$ and  at large $\ell \theta \gg 1$}, we have $P_{\ell}(\cos\theta)\propto 1/\sqrt{\ell \theta} \cos{(\ell \theta -\pi/4)}+\O((\ell\theta )^{-3/2})$. As a result, 
 the peak of $\ell(\ell+1)  \ol{C}^\F_{\ell}$ is roughly located at $\ell \simeq \ell_p \simeq 1/P_{\rm DW} R$. Thus, we obtain $\ell(\ell+1)  \ol{C}^\F_{\ell} \propto \beta^2 (\ell/\kappa)^{{2}}$ for $\ell < \ell_p$. This explains the $\ell$-dependence as well as the reason why the peak height is insensitive to  $\kappa$.
We emphasize again that $\ol{C}_\ell^\F \propto \beta^2$ and thus the isotropic and anisotropic CBs are correlated. 
The characteristic angular power spectrum correlated with the isotropic CB is peculiar to our KBCB scenario.}

\begin{figure}[!t]
\begin{center}  
   \includegraphics[width=145mm]{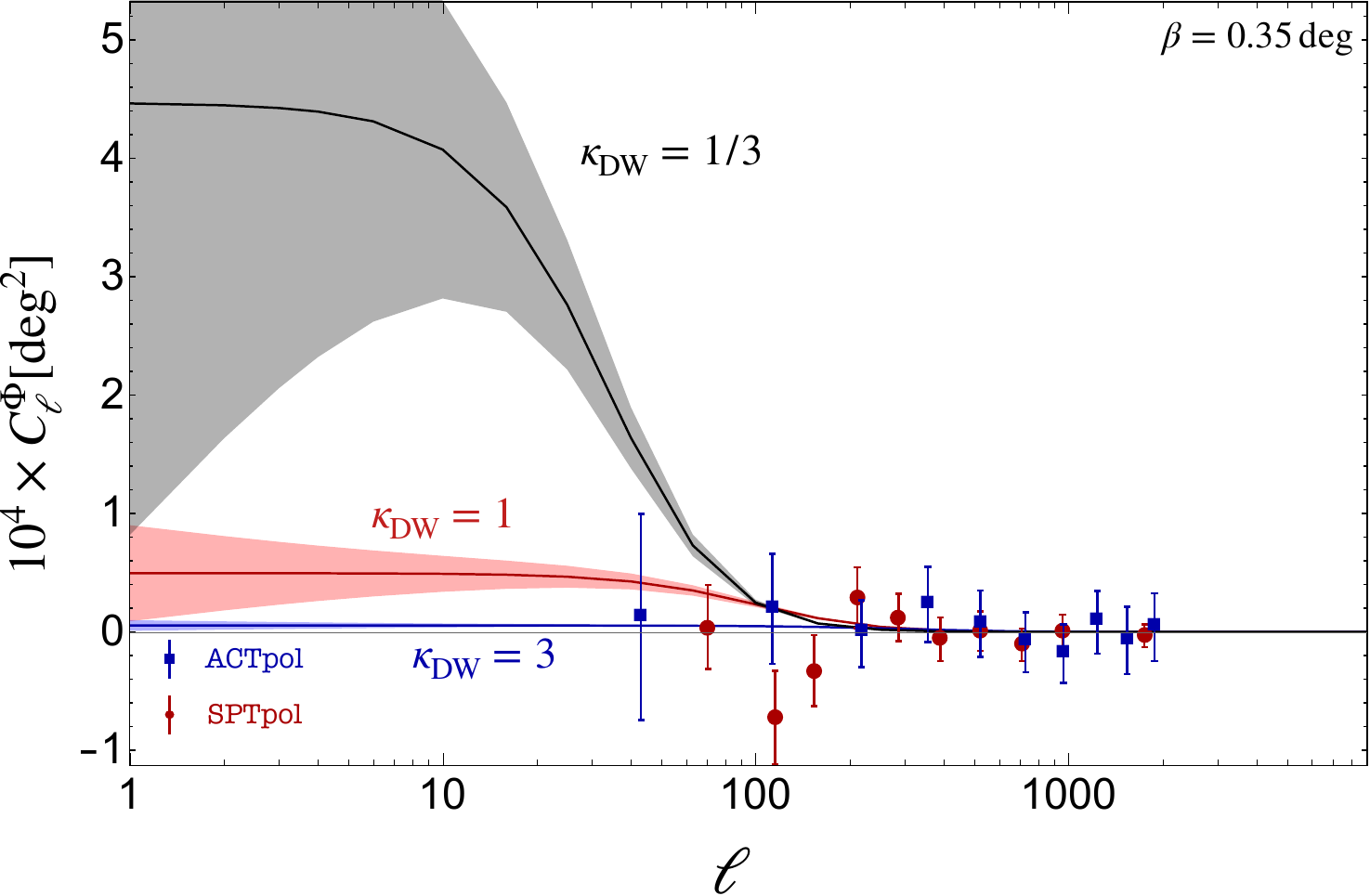}
      \end{center}
\caption{{
Same as Fig.~\ref{fig:Cl0}, but plotted with a different scale to facilitate comparison with the observed data of  ACTpol~\cite{Namikawa:2020ffr} and SPTpol~\cite{Bianchini:2020osu}. Those data points are  adopted from Ref.\,\cite{Bianchini:2020osu}.}
} \label{fig:Cl}
\end{figure}

{The question is whether the predicted anisotropic CB in our scenario is consistent with current observations. 
In Fig.~\ref{fig:Cl}, we display the comparison between the predicted $C_\ell^\F$ and the data obtained by the ACTpol and SPTpol~\cite{Bianchini:2020osu, Namikawa:2020ffr} in blue and red points, respectively.  
We expect that, for $\kappa_{\rm DW}(0.35{\rm deg}/\beta)\lesssim 1$, the predicted $C_\ell^\F$ is sizable 
 at low $\ell \lesssim 100$, and may be tested in the future observation of anisotropic CB.
Further observations of both isotropic and anisotropic CB satisfying the above features will be a smoking-gun evidence for our KBCB.}

\begin{figure}[!t]
\begin{center}  
   \includegraphics[width=145mm]{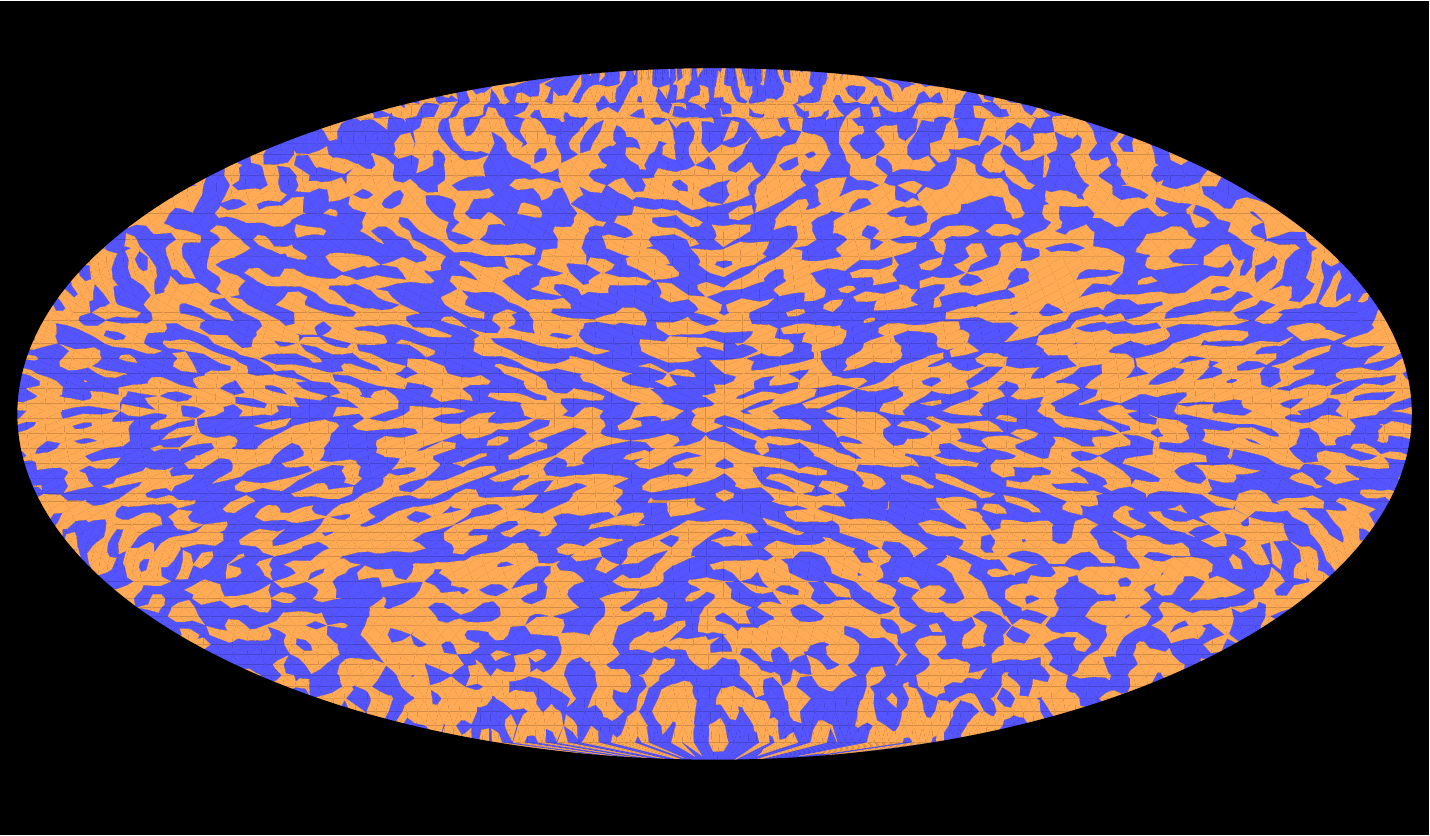}
      \end{center}
\caption{
An example of the polarization map of $\F(\Omega)$ for photons where in the orange (blue) region we have $\Delta \F= {\rm sign}[\f_{\rm today}-\f_{\rm LSS}(\Omega) ] c_\gamma \alpha $ ($\Delta \F= 0$).  } \label{fig:CMB}
\end{figure}

The actual angular dependence of $\Delta \Phi$ in {a single sample} can be obtained by a dedicated analysis based on the lattice simulation of the domain wall formation. We leave it for the future work, and instead, we have made a
mock sample by randomly generating domains with its typical correlation length of order the Hubble horizon at the recombination. 
See Fig.\,\ref{fig:CMB}, where one can see that  $\Delta \Phi$ takes discrete values {in each domain.}
We have also confirmed that the anisotropic birefringence induced by such a mock sample satisfies the above mentioned features as well as the current bound~\cite{Namikawa:2020ffr,Bianchini:2020osu} {by performing the multipole expansion.}

To be more {realistic}, the boundary between domains should be blurred to some extent, because
the LSS has a finite depth of order $0.1 H^{-1}_{\rm LSS}$  and the domain walls also move at a finite speed, probably
close to the speed of light. How much the boundary is blurred depends on the angle the domain wall makes with the LSS and the velocity.
However, this boundary effect does not change our estimate on the isotropic birefringence, while it slightly affects the anisotropic one at $\ell \gtrsim \ell_p$. To make a precise estimate, more detailed and dedicated analysis is warranted.

Lastly, let us mention
a possibility of domain walls passing through the Earth. If we live in the vacuum R,  $\Delta \Phi$  becomes smaller by $c_\gamma \alpha$
after the domain wall goes through the Earth.
Also, if the ALP is  coupled to the SM fermions, a distinctive signal pattern may be obtained in detectors located at different places in   the GNOME experiment~\cite{Masia-Roig:2019hsy}. That said,
 the probability of this event to take place in a period of $\D t$ is  $\sim \D t\, H_0$  with $H_0$ being the Hubble constant, and it is very unlikely to happen in the data-taking time, as long as the domain walls follow the scaling solution.
 Also the domain wall may spend $\Delta t\sim 10^3 s \(\frac{10^{-18}\EV}{m_\phi}\)$ to go through the Earth, which may also be too long to observe~\cite{Masia-Roig:2019hsy}.

\section{Discussion and conclusions }

\label{sec:dis}

In our scenario, domain walls separating the adjacent vacua are formed without strings. Therefore, the winding number is trivial everywhere in the  universe. This is the reason why the domain walls along the line of sight are irrelevant for the net rotation angle of the CMB polarization. This also explains why our scenario works even if the vacua $L$ and $R$ are physically identical. 

{
Our mechanism of the domain wall formation works in a wide class of potentials with degenerate vacua, which will
lead to the KBCB. 
So far we have assumed $\sigma_\theta \lesssim 2\pi$, but even if $\sigma_\theta \gtrsim 2\pi$, domain walls without strings satisfying the scaling solution are considered to be formed. 
This is because the domain walls quickly evolve with time after the formation, and we expect that only those vacua satisfying the criterion due to the percolation theory will remain. In other words, domains near the edge of the distribution will soon disappear due to the domain-wall dynamics, and so, they are not relevant. 
Then, we will have a similar KBCB in this case.
A possible difference is that $\Delta\Phi$ may take more than two discrete values, but it requires detailed numerical simulations how the domain wall network evolves in this case, 
which is  beyond the scope of this paper.
In general, as the number of degenerate vacua increases, the anisotropic CB can be more significant compared to the isotropic one. This is because the distribution of the vacua on the sphere of a radius $r$ around us can be understood in terms of the random walk from us at $r=0$, and we may happen to  live in the vacuum close to the center of the probability distribution.
Our argument {on the CB} can be straightforwardly extended to such cases.  However, when the potential has a local false vacuum, the discussion may become more complicated than what we have described. 
}

{
One of the central assumptions in the above discussion is the scaling solution of domain walls. It was argued in Refs.~\cite{Coulson:1995nv,Larsson:1996sp,Hindmarsh:1996xv,Correia:2014kqa,Correia:2018tty} that, 
with a biased initial condition, the domain-wall network follows the scaling solution only for a finite time. In the case of a $Z_2$ domain-wall model, the domain walls collapse during several Hubble times after the formation unless the probability falling into one of the vacua is very close to $0.5$. However, this conclusion depends not only on the initial probability distribution, but also on the (implicitly) assumed power spectrum. 
In particular, in these studies, the initial fluctuation of the scalar field is taken to be a white noise, and there are no fluctuations at superhorizon scales. In the presence of fluctuations at superhorizon scales, one of the vacua cannot be chosen over the entire universe, because the averaged
scalar field over a Hubble horizon can be either positive or negative depending 
on the superhorion modes.\footnote{We thank Naoya Kitajima for checking this with his numerical code.} 
For a more quantitative study on the domain-wall evolution, we need dedicated calculations,
which will be given elsewhere. Here we simply mention that the lifetime of the 
ALP domain wall network can be much longer than the cases studied in the above references if the initial fluctuations are generated during inflation.
On the contrary, in the case that the initial ALP fluctuation is induced by the PQ phase transition for the QCD axion, the lifetime could be finite, in which case our results will be limited to
ALP masses close to the bound \eq{preCMB}.\footnote{{If the distribution of $\phi$ is much broader than $f_\phi$,  the lifetime of domain walls might be prolonged.}}
}

{It is also possible that domain walls are unstable and decay, if the degeneracy of the vacua is lifted by another shift symmetry breaking term. If the domain walls disappear after recombination, the bound (\ref{tension}) from the cosmological domain wall problem will be significantly relaxed, and the viable parameter space might be enlarged. }

So far we have discussed the domain wall formation before the recombination.
The domain wall formation may also happen after the recombination, in which case the typical curvature of the axion potential should be lighter than $10^{-29}\EV.$
In this case, we have an almost isotropic polarization without the KB structure.\footnote{
If the axion potential has a local false vacuum as in the axion monodromy, a completely isotropic CB may be generated; the axion is trapped in the false vacuum and undergoes tunneling after phase transition. In this case the prediction may be  gravitational waves generated when the bubble walls collide.}
From all directions the polarization angle changes by a value corresponding to the difference of $\vev{\phi}/f_\phi$ at our vacuum and the value, $\phi_{\rm LSS}/f_\phi$, at the last scattering. 
The anisotropic CB is also generated since the ALP distribution has a variance of $\D\phi_{\rm LSS}/f_\phi\sim H_{\rm inf}/f_\phi^{\rm (inf)}$. 
Most predictions in this case, however, will be similar to that from the slow-rolling ALP scenario~\cite{Carroll:1998zi,Lue:1998mq, Minami:2020odp, Fujita:2020ecn}. The difference is the presence of  domain walls today which may be searched for by other means.

We have assumed that the domain walls follow the scaling solution, but if the domain walls are formed during inflation,
the number density of walls is exponentially suppressed, and it does not reach the scaling solution. If the domain wall formation (or the spontaneous breaking of the U(1)$_{\rm PQ}$ symmetry) takes place around the e-folding number about $50-60$, 
it is possible that domain walls enter the horizon relatively recently, well after the recombination. In particular, domain walls may be bounded by strings in this case. 
We expect both isotropic and anisotropic CB of the same order as before, but the anisotropy exists only at large scales corresponding to the typical size of the domain walls. 

An interesting question is whether the axion can be the dominant dark matter since the mass heavier than $\O(10^{-22}\EV)$ is allowed~\cite{Hui:2016ltb,Marsh:2018zyw,Irsic:2019iff,Schutz:2020jox}. However, to explain the dark matter abundance from the misalignment mechanism,
{ 
\beq 
\Omega_{\phi}^{\rm mis} h^2 \sim 10^{-14} 
\bigg(\frac{m_\phi}{10^{-22}\,\text{eV}}\bigg)^{1/2} 
\bigg(\frac{f_\phi}{10^{10}\,\text{GeV}}\bigg)^2
\eeq
by taking the initial misalignment angle $\sim \pi$},
the required decay constant is too large to be consistent with the  viable parameter region for $c_\g=\O(1)$ (See Fig.~\ref{fig:region}). We would also have the severe isocurvature bound if the fluctuation of the ALP is originated from the quantum fluctuation during inflation. 
Thus, 
we have to say it is difficult to make the ALP explain all dark matter. 
Also, axions are continuously generated by domain walls which follow the scaling solution, its abundance is much smaller than the dark matter abundance for domain walls satisfying (\ref{tension}). {That said, the subdominant ALP DM is the prediction of our scenario. The detection of it may be an interesting future experimental approach (the current proposals are, however, difficult to constrain it~\cite{Fedderke:2019ajk, Fujita:2018zaj,Ivanov:2018byi,Caputo:2019tms,Fedderke:2019ajk}.)}

Instead, as we have discussed, the PQ fermion {or the QCD axion relevant to the ALP domain wall formation} is a good candidate of dark matter.

\subsection*{Conclusions}
In this paper, we have proposed {simple mechanisms} for the ALP domain wall formation without strings, where domain walls separate the two adjacent vacua whose existence is generally expected from the discrete shift symmetry \eq{shift}. The condition for the domain wall formation can be naturally satisfied if the PQ scalar has a non-minimal coupling to gravity {or if the ALP mixes another axion such as the QCD axion.}
The domain wall, if formed before the recombination, both isotropic and anisotropic CB are predicted. Interestingly, the former agrees with the recently reported value \eq{measure}, and the latter should reflect the domain wall configuration at the LSS, independent of the domain walls along the line of sight. We stress that the isotropic CB is due to the spontaneous breaking of the exchange symmetry of the two vacua; we must live in one of the two vacua. The detection of the two different CB with the peculiar features will be a smoking-gun evidence for our scenario.
Further observation and analysis of the CMB polarization may reveal the information of order KB encoded on the LSS. 

\appendix
\section*{Appendix A: cosmic variance}
{Let us make a simple order estimate on the cosmic variance. The definition of $C^\F_l$ (without taking the ensemble average) is given as 
\beq
\laq{def1}
C_{\ell}^\F=\frac{1}{4\pi} \int{d^2 \Omega_1 d^2 \Omega_2} P_{\ell}(\hat{\Omega}_1\cdot \hat{\Omega}_2) \D \tl{\F} (\Omega_1)\D \tl{\F} (\Omega_2).
\eeq
Then we obtain the cosmic variance with 
\beq
\vev{\frac{(C_{\ell}^\F-\ol{C}_\ell^\F)^2}{(\ol{C}_\ell^\F)^2}}=\frac{\vev{\(C_{\ell}^\F\)^2}-\(\ol{C}_\ell^\F\)^2}{(\ol{C}_\ell^\F)^2}
\eeq
The first term on the numerator is the only term that we should consider carefully. 
According to \eq{def1} this is 
\beq
\vev{\(C_\ell^\F\)^2}=\frac{1}{(4\pi)^2} \int{
d^8 \Omega 
P_{\ell}(\hat{\Omega}_1\cdot \hat{\Omega}_2) P_{\ell}(\hat{\Omega}_3\cdot \hat{\Omega}_4) \vev{\D \tl{\F} (\Omega_1)\D \tl{\F} (\Omega_2)\D \tl{\F} (\Omega_3)\D \tl{\F} (\Omega_4)}.}
\eeq
Here we assume that if $\hat{\Omega}_i \cdot \hat{\Omega}_j\ll 1-1/(P_{\rm DW} R)^2$, i.e. the positions for the angular coordinates have a physical distance much larger than the horizon size, 
the polarization are statistically independent. Then we can separate the integral regime as 
\begin{align}
\int  =
&\int_{(\Omega_1 \Omega_2)(\Omega_3\Omega_4)}+\int_{(\Omega_1 \Omega_3)(\Omega_2\Omega_4)}+\int_{(\Omega_1 \Omega_4)(\Omega_2\Omega_3)}\laq{22}\\ 
&+\int_{(\Omega_1 \Omega_2 \Omega_3\Omega_4)}\laq{4}
\end{align}
Here the positions for angular coordinates in side the bracket, $()$, have distance with each other within $\O(1\text{-}10)$ Hubble horizon size. We have dropped the integral with $(\Omega_i)$ since it vanishes according to our assumption, e.g.
\begin{align}
 &\int_{(\Omega_1)(\Omega_2)(\Omega_3)(\Omega_4)}{d^8 \Omega} P_{\ell}(\hat{\Omega}_1\cdot \hat{\Omega}_2) P_{\ell}(\hat{\Omega}_3\cdot \hat{\Omega}_4) \vev{\D \tl{\F} (\Omega_1)\D \tl{\F} (\Omega_2)\D \tl{\F} (\Omega_3)\D \tl{\F} (\Omega_4)}\\
&=\int_{(\Omega_1)(\Omega_2)(\Omega_3)(\Omega_4)}{d^8 \Omega} P_{\ell}(\hat{\Omega}_1\cdot \hat{\Omega}_2) P_{\ell}(\hat{\Omega}_3\cdot \hat{\Omega}_4) \vev{\D \tl{\F} (\Omega_1)}\vev{\D \tl{\F} (\Omega_2)}\vev{\D \tl{\F} (\Omega_3)}\vev{\D \tl{\F} (\Omega_4)}\\
&= 0.
\end{align} 
The corresponding terms to \eq{22} and \eq{4} in general do not vanish. However \eq{22} and \eq{4} have different  phase spaces of $\O(1/(P_{\rm DW} R)^4)$ and $\O(1/(P_{\rm DW} R)^6)$ respectively. Therefore the dominant contribution should be from \Eq{22} if the integrants for \Eqs{22} and \eq{4} do not have large hierarchy.  
In particular \eq{22} has a similar contribution to the usual Gaussian distribution case. 
This can be found from explicitly calculations, 
\begin{align}
&\int_{(\Omega_1 \Omega_2)(\Omega_3 \Omega_4)}{d^8 \Omega} P_{\ell}(\hat{\Omega}_1\cdot \hat{\Omega}_2) P_{\ell}(\hat{\Omega}_3\cdot \hat{\Omega}_4) \vev{\D \tl{\F} (\Omega_1)\D \tl{\F} (\Omega_2)\D \tl{\F} (\Omega_3)\D \tl{\F} (\Omega_4)}\\ 
&=\int_{(\Omega_1 \Omega_2)(\Omega_3 \Omega_4)}{d^8 \Omega} P_{\ell}(\hat{\Omega}_1\cdot \hat{\Omega}_2) P_{\ell}(\hat{\Omega}_3\cdot \hat{\Omega}_4) \vev{\D \tl{\F} (\Omega_1)\D \tl{\F} (\Omega_2)}\vev{\D \tl{\F} (\Omega_3)\D \tl{\F} (\Omega_4)}\\
&=\(\int- \int_{(\Omega_1\Omega_2\Omega_3\Omega_4)}\){d^8 \Omega} P_{\ell}(\hat{\Omega}_1\cdot \hat{\Omega}_2) P_{\ell}(\hat{\Omega}_3\cdot \hat{\Omega}_4) \vev{\D \tl{\F} (\Omega_1)\D \tl{\F} (\Omega_2)}\vev{\D \tl{\F} (\Omega_3)\D \tl{\F} (\Omega_4)}\laq{bypart}\\
&= \frac{(4\pi)^2}{(2\ell +1)^2}  \sum_{m' m}\vev{a_{l-m} a_{l m}} {\vev{a_{l-m'} a_{l m'}}} +\O(1/(P_{\rm DW} R)^6)\\
&\simeq  (4\pi)^2  \(\ol{C}_\ell^\F\)^2
\end{align}
where we used \Eq{22} in the second equation and again neglected the terms with $\vev{\D\tl{\F}(\Omega_i)}$; we have used $P_{\ell}{(\hat{\Omega}\cdot \hat{\Omega}')}= \frac{4\pi}{2\ell+1}\sum_{m}Y_{\ell m} (\Omega )Y_{\ell -m}(\Omega')$ and neglect the higher order term in the last equation. 
Similarly we obtain 
\begin{align}
&\int_{(\Omega_1 \Omega_3)(\Omega_2 \Omega_4)}{d^8 \Omega} P_{\ell}(\hat{\Omega}_1\cdot \hat{\Omega}_2) P_{\ell}(\hat{\Omega}_3\cdot \hat{\Omega}_4) \vev{\D \tl{\F} (\Omega_1)\D \tl{\F} (\Omega_2)\D \tl{\F} (\Omega_3)\D \tl{\F} (\Omega_4)}\\
&= \frac{(4\pi)^2}{(2\ell +1)^2}  \sum_{m' m}\vev{a_{l-m} a_{l -m'}} {\vev{a_{l m} a_{l m'}}} +\O(1/(P_{\rm DW} R)^6)\\
&\simeq  \frac{(4\pi)^2}{(2\ell +1)} \(\ol{C}_\ell^\F\)^2.
\end{align}
By performing the last integral in \eq{22} we get a same result. 
Then in total we get 
\beq 
-\(\ol{C}_\ell^\F\)^2+\vev{\(C_\ell^\F\)^2}= \frac{2}{(2\ell+1)} \(\ol{C}^\F_{\ell}\)^2+\O(\(P_{\rm DW}R\)^{-6}).
\eeq 
At the peak we expect $\ell_p\sim 1/(P_{\rm DW}R)\sim \O(100),$ and $\ol{C}_{\ell}^\F\sim \beta^2/(P_{\rm DW} R)^2$ {at $\ell < \ell_p$} as can be seen from Fig.\,\ref{fig:Cl}.  Consequently if  $\ell$ is not {much} larger than $\ell_{p}$, the cosmic variance can be well approximated as \beq \D C^\F_\ell\simeq \sqrt{\frac{2}{2\ell+1}}C^\F_\ell.\eeq 
However if $\ell \gg \ell_p$, by noting $\ol{C}_{\ell}^\F < 1/\ell^2$ from Fig.~\ref{fig:Cl0}, we might have regime that the $\O((P_{\rm DW} R)^{-6})$ term becomes important if it decrease slower than $(\ol{C}_\ell^\F)^2/(2\ell+1)$ does. 
{On the other hand, our simple model is considered to receive larger corrections, 
because we need to take account of the finite width of the LSS, and so on.}
}

\section*{Acknowledgments}
F.T. thanks Diego Gonzalez, Naoya Kitajima, and Masaki Yamada for fruitful discussion
on the possibility of using the domain walls to explain the cosmic birefringence. 
F.T. was supported by JSPS KAKENHI Grant Numbers 17H02878, 20H01894, 20H05851, 
and by World Premier International Research Center Initiative (WPI Initiative), MEXT, Japan. 
W.Y. was supported by JSPS KAKENHI Grant Number 19H05810.

\end{document}